# Mining Multi-level Frequent Itemsets under Constraints

Mohamed Salah GOUIDER

*BESTMOD Laboratory*
*Institut Supérieur de Gestion*
*41, rue de la liberté,*
*cite Bouchoucha*
*Bardo, 2000, Tunis, Tunisia*
*E-mail: ms.gouider@isg.rnu.tn*

Amine FARHAT

*BESTMOD Laboratory*
*Institut Supérieur de Gestion*
*41, rue de la liberté,*
*cite Bouchoucha*
*Bardo, 2000, Tunis, Tunisia*
*E-mail: farhat_amine@yahoo.fr*

*ABSTRACT*

*Mining association rules is a task of data mining, which extracts knowledge in the form of significant implication relation of useful items (objects) from a database. Mining multilevel association rules uses concept hierarchies, also called taxonomies and defined as relations of type 'is-a' between objects, to extract rules that items belong to different levels of abstraction. These rules are more useful, more refined and more interpretable by the user. Several algorithms have been proposed in the literature to discover the multilevel association rules. In this article, we are interested in the problem of discovering multi-level frequent itemsets under constraints, involving the user in the research process. We proposed a technique for modeling and interpretation of constraints in a context of use of concept hierarchies. Three approaches for discovering multi-level frequent itemsets under constraints were proposed and discussed: Basic approach, "Test and Generate" approach and Pruning based Approach.*

## 1. Introduction

The Knowledge Discovery from Database (KDD), means the non-trivial process of identifying, from the data, patterns or valid knowledge which is new, useful and understandable [1]. The KDD is motivated by the huge volumes of data collected around the world, and the more efficient and reliable environment of exchange of data provided by systems and networks. Data mining is the core step of KDD process, defined as the set of intelligent, complex and highly sophisticated data processing techniques, used to extract knowledge. Knowledge can take several forms depending on the purpose of the user and the data mining algorithm. Mining association rules is a data mining task which consists in extracting meaningful relationships of the form (X implies Y) between objects (Items) of a database, such as X and Y are subsets of items. The validity of an association rule is defined by two measures where the threshold is defined by user: the first measure is the support which means the scope of the rule, i.e. the frequency of the set (X UNION Y) in the database. The second measure is the confidence that means the accuracy of the rule, i.e. the conditional probability of occurrence of Y knowing X. This problem was proposed in [2] for the analysis of transactions of a sale database. Since then, mining association rules has become a very important task of data mining and has demonstrated efficacy in diverse application areas: telecommunications, medical diagnostics, space exploration ... This problem has been addressed in several articles in the literature [3, 4, 5, 6, 7], which allowed the development of several algorithms for discovering association rules.

Approaches mentioned above are aimed at discovering association rules at the terminal level of abstraction, i.e. the association rules containing only the items belonging to the transactions of database. However, there is a need in many applications to association rules at higher levels of abstraction, these are multi-level association rules [9] or generalized association rules [8]. Mining multi-level association rules is motivated by several reasons, such as:

- Association rules at the lowest level of abstraction may not satisfy the support constraint. Thus, one may omit several rules potentially useful.
- The multi-level association rules are more refined, give a global view and are more interpretable and more understandable to the user.
- The multi-level association rules can provide solutions to the problem of redundant or unnecessary rules, often encountered in real world applications.

To extract multi-level association rules, concept hierarchies or items taxonomies are needed. A concept hierarchy is modeled by a directed acyclic graph (DAG) whose nodes represent items and arcs represent 'is-a' relations between two items. Concept hierarchies represent the relationships of generalization and specification between the items, and classify them at several levels of abstraction. These concept hierarchies are available, or generated by experts in the field of application. Figure 1 illustrates an example of a concept hierarchy on food products. The problem of discovering multi-level association rules has been treated in several articles in the literature that suggested many methods for solving the problem. Many studies are focused on the problem of finding multi-level frequent itemsets, which represent the main and most complex stage in the process of extracting association rules.

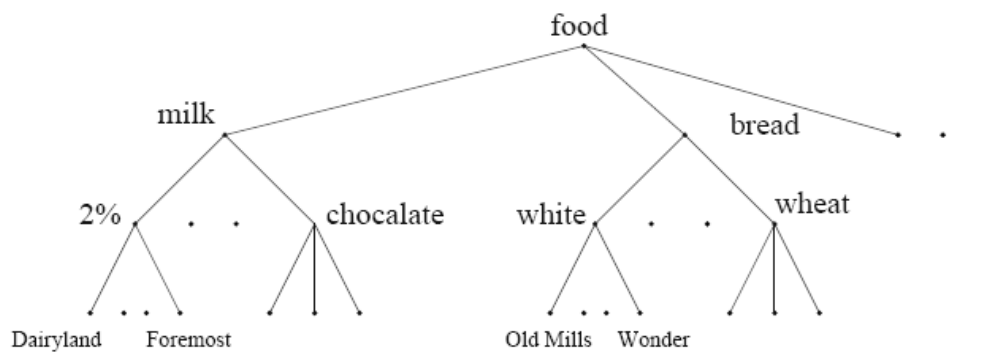

**Figure 1. Example of a concept hierarchy on food products**

This paper deals with the problem of finding frequent multi-level itemsets under constraints, given a deep belief of the importance of involving the user in the process of mining association rules. This process leads to developing more reliable and efficient solutions, especially for processing large volumes of data.

Our contribution is to propose, first, a technique of definition and interpretation of constraints in the context of use of concept hierarchies. Then, three approaches and algorithms for discovering frequent multi-level itemsets under constraints are proposed and discussed: basic approach, approach 'test and generate' and pruning based approach. To develop the approach by pruning, some changes will be made to the technique of definition and interpretation of the constraints.

This article is organized as follows. The problem of mining multi-level association rules is specified in details in section 2. Main approaches proposed in the literature will be presented in the same section. Section 3 is devoted to solving the problem of mining frequent multi-level itemsets under constraints. The conclusion and the perspectives of this work are presented in section 4.

## 2. Mining multi-level association rules

At this section, a specification of the problem and a brief description of the techniques of extracting multi-level association rules proposed in the literature, are presented.

### 2.1. Problem specification

Let I = {$i_1$, $i_2$, ... ... ., $i_m$}, a set of m items, also called literals. Let T = {$t_1$, $t_2$, ... ...., tn}, a database of n transactions, each transaction $t_i$ is composed of a unique identifier (TID) and a subset i of I, i is composed of k items and i is defined as a k-itemset. HC is a concept hierarchy on the items of I, HC is also called taxonomy. It is modeled by a directed acyclic graph. An arc of HC represents an "is-a" relationship between the source and the destination. A node refers to an item of I. Let p and c, two nodes of HC, and there is an arc from p to c, p is said parent and c is said son. An item is not the parent of itself since the graph is acyclic. Transactions T contain only the items belonging to the lowest level (Terminal level). In taxonomy, levels are numbered from 0, as the level 0 represents the level Root. Items belonging to a level *l*, are numbered with respect to their parent in an ascending order, this coding was proposed in [9] for reasons of simplification. In Figure 1, the item milk, for example, takes the code 1**, since it belongs to level 1, the Dairyland item (terminal item) takes the code 111, which gives clear information about its position in the hierarchy and its parents.

The problem of mining multi-level association rules is to find the association rules containing items belonging to the different levels of abstraction, meeting the minimum thresholds of support and confidence, knowing that a transaction t supports an item x if and only if, x belongs to t or x is a parent of an item belonging to t. Similarly to the methods of discovering single-level association rules, multi-level association rules algorithms are mainly based on the discovery of frequent itemsets.

### 2.2. Algorithms for mining multi-level association rules: An Overview

As indicated earlier, the problem of mining multi-level association rules has been covered in several researches. In [9], the authors have proposed series of algorithms for discovering multi-level frequent itemsets: ML-T2, ML-T1, ML-TMax, ML-T2+…. All these algorithms implement a top-down deepening method that starts with the treatment of the highest level of abstraction and then the lowest levels. These algorithms use different minimum support thresholds for the different levels of abstraction, these thresholds values decrease in the hierarchy of concepts for the following reasons:

- Avoid the generation of unnecessary or obvious association rules at high abstraction levels.
- Avoid the omission of useful association rules at low abstraction levels.

Algorithms proposed in [9] can generate frequent itemsets for each level independently. In [8], a new approach to solve the problem has been proposed. This approach consists in generating, in a first step, an extended version of the database, so that each transaction gives rise to a new transaction which, in addition to the initial items, contains the ancestors of each item of the transaction. Then, all transactions will contain items from different levels of abstraction in the concept hierarchy. In a second step, algorithms for discovering frequent itemsets are applied on this new extended version of database. Indeed, this approach can generate frequent itemsets containing, simultaneously, items belonging to several levels of abstraction. Three algorithms have been proposed, implementing several methods of optimization and performance improvement: Basic, Cumulate, Stratification. In [10], the *PRUTAX* algorithm was proposed for mining multiple level frequent itemsets, implementing a hash tree based method in order to reduce the number of support calculations as it counts only the supports for candidate itemsets whose ancestors are all frequent.

In [11], the authors have reviewed the proposed algorithms in [9], and proposed new improved and optimized algorithms: *ML-T2L1, ML-T1LA, ML-TML1, ML-TML1, ML-T2LA*. In [12], a formal framework for generalized itemsets was defined based on two relationships: Superset-Subset and Parent-Child. Then, the *SET* algorithm was proposed to enumerate all generalized frequents itemsets, *SET* uses two constraints based on the defined relationships on itemsets in order to avoid support calculations for infrequent itemsets. In [13], a top-down progressive deepening method of mining cross level frequent itemsets has been proposed, i.e. in what appear simultaneously items from different levels of abstraction. Their method, based mainly on the work of Han and Fu [9, 11], is to create a data structure that combines incrementally the 1-itemsets (items) for each level of abstraction. This structure is used to generate 2-itemsets candidates for all levels of abstraction, which are cross-level itemsets (containing items from several levels of abstraction and not just the level being processed). This type of frequent itemsets and association rules can reveal new correlations potentially more useful for the user.

## 3. Algorithms for Mining Frequent Multi-level Itemsets under Constraints

The objective of this paper is to develop a method of finding frequent multi-level itemsets under constraints. Our method is to consider the needs of the expert (user), and to give him the possibility to manage and personalize the research process. To achieve this goal, in the beginning the technique developed for modeling the constraints in a context of use of concept hierarchies on the items of the database is presented. Then, scenarios considered and studied to solve the problem will be presented in details.

### 3.1. Modeling the constraints of existence on association rules

The constraints on the association rules are the criteria defined by the user to customize and guide the search process to better achieve its objectives. The support and confidence are two fundamental constraints in the process of discovering association rules. An association rule is not accepted if it does not meet these two constraints. Particular interest in this work is restricted to the constraints of existence, which enables the user to filter the items, which may be included in the itemsets to discover. In [14], the authors proposed a technique for modeling constraints that can be integrated into the search process of frequent itemsets. This technique uses the principles of classic logic. It considers an existence constraint as a Boolean expression in disjunctive normal form: a disjunction of conjunctions, treating an item as a literal. To clarify this technique, we give the definitions of some concepts from classical logic:

- **A literal:** a literal is an atom (also called positive literal) or the negation of an atom. The atoms form clauses.

- **The conjunction or AND logic:** is a logical operator in the calculation of the proposals. The proposition obtained by linking two propositions by this operator is also called logical product. The conjunction of two propositions P and Q is true if both propositions are simultaneously true, otherwise it is false. The conjunction is: P AND Q.

- **A disjunction or OR logic:** is a logical operator in the calculation of the proposals. The proposition obtained by linking two propositions by this operator is also called logical sum. The disjunction of two propositions P and Q is true when one of them is true and is false when both are simultaneously false. The disjunction is: P OR Q.

- **Disjunctive Normal Form:** a disjunctive normal form (DNF) is a standardization of a logical expression which is a disjunction of conjunctive clauses.

**Specification of this technique:**

According to this technique and based on the definitions above, a constraint CT will be structured as follows:

$$CT = c_1 \text{ OR } c_2 \text{ OR } c_3 \text{ OR} \ldots \ldots \ldots \ldots \text{OR } c_n$$

Each $c_i$ is a conjunction with the following structure:

$$C_i = e_{i1} \text{ AND } e_{i2} \text{ AND } e_{i3} \ldots \ldots \ldots \ldots \ldots \text{AND } e_{ini}$$

An element $e_{ij}$ represents the elementary level; it must have valid logical value (True or False). It consists of a literal (Item, in our case) and, if necessary, a sign of negation. To be valid and satisfies a constraint, an itemset must have the logical value 'True' for at least one conjunction $C_i$ of CT.

**Example:**
Given the following items: A, B, C, D, and E.
And a constraint CT01:
*CT01 = (A AND B) OR ((NOT A) AND D) OR (D AND C).*

To be valid for the constraint CT01, an itemset it01 must satisfy at least one of the following clauses:
- A and B, in it01.
- Not A and D in it01.
- D and C, in it01.

It01 may, for example, be one of the following:
- it01 = {A, B, C}
- it01= {D, C, A}

Relying on this technique of modeling constraints, several algorithms for discovering frequent itemsets satisfying the constraints of existence (Constraints controlling the appearance of items in itemsets), defined by the user, have been proposed in [14].

### 3.2. Modeling the constraints of existence in a context of use of concept hierarchies

In this paper, the modeling technique of constraints proposed in [14] and presented in Section 3.1 is extended to make it applicable in the context of multi-level frequent itemsets. In a context of use of concepts hierarchies, a constraint CT keeps the same structure as:

$$CT = c_1 \text{ OR } c_2 \text{ OR } c_3 \text{ OR} \ldots \ldots \ldots \ldots \text{OR } c_n$$

Each $C_i$ is a conjunction with the following structure:

$$C_i = e_{i1} \text{ AND } e_{i2} \text{ AND } e_{i3} \ldots \ldots \ldots \ldots \ldots \text{AND } e_{ini}$$

The difference compared to the technique proposed in [14], is that the element $e_{ij}$ may be composed of an item belonging to different levels of the concept hierarchy, and not only to the terminal level. This influences the way of interpretation of the constraint itself. The interpretation of a constraint in the context of use of concept hierarchies is defined as follows:

Consider the structure of the constraint CT, illustrated above.
Let $e_{ij}$, a basic element in one of the conjunctions ($C_i$) of constraint CT. Two scenarios may arise:

1. $e_{ij} = I01$, as I01 is an item:

   For an itemsets satisfies IT01 element $e_{ij}$, I01 must contain IT01 or IT01 contain at least one of the descendants of I01 in the concept hierarchy.

2. $e_{ij}$=(NOT I01), as I01 is an item:

For an itemsets satisfies IT01 element $e_{ij}$, IT01 mustn't contain I01, or any item from the descendants of I01.

The principle of modeling and interpretation of constraints is implemented in the following algorithms.

### 3.3. Algorithms for mining frequent multi-level itemsets under constraints

The existing algorithms for discovering multilevel frequent itemsets do not address the problem of constraints which can be defined by the user to customize the results. In this paper, constraints of existence are particularly treated. As part of this goal, first and foremost, a technique of modeling constraints in a context of use of concept hierarchies is developed, by extension of the technique proposed in [14], (see Section 3.2). This technique allows easier and deterministic integration of constraints in the algorithms. In this section, a detailed study of different scenarios for solving the problem is presented, as well as the developed algorithms. These are based on the algorithms for discovering multi-level frequent itemsets proposed in [9] and [11]. An example of a concept hierarchy, with fictitious items for reasons of simplification and interpretation, is presented in Figure 2. An example of a database used for the illustration of the application of algorithms, is presented in Table 1. The concept hierarchy in Figure 2 consists of 3 levels of abstraction, codification of items related to their position in the hierarchy is used [9].

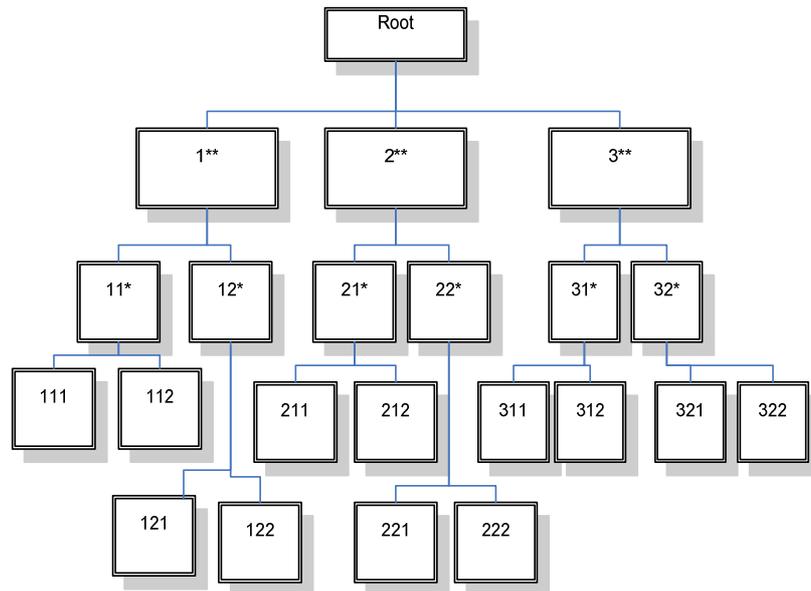

**Figure 2. The Concept Hierarchy**

**Table 1: Database**

| TID | Items |
|---|---|
| 1 | 111, 212, 221, 312, 321 |
| 2 | 111, 122, 312, 321, 222, 212 |
| 3 | 321, 322, 122, 112, 212 |
| 4 | 212, 111, 122, 312, 322, 211 |
| 5 | 111, 211, 221, 321 |
| 6 | 321, 211, 121, 122 |
| 7 | 111, 212, 311, 321 |
| 8 | 212, 112, 122, 322, 211 |

**3.3.1. Scenario 1: Basic Algorithm :** The first scenario considered in the discovery of multi-level frequent itemsets under constraints proceeds as follows:
- Search for frequent itemsets for each level independently, based on the Apriori algorithm and using different values of the support: A minimum value of support for each level is defined.
- After finding the frequent itemsets of a level l, the validity of these itemsets against the constraint defined by the user is verified.
- Elimination of frequent itemsets which do not satisfy the constraint.

This approach is the solution "Generate and test". It proceeds to verify the satisfaction of constraints after the discovery of frequent itemsets. This approach is presented in the following algorithm in table 2:

**Table 2: Pseudo-code of the first algorithm of extraction of multi-level frequent itemsets under constraints: Scenario 1.**

```
Algorithm 1: Basic Version:
Input: T: Data base transactions
       HC: Hierarchy of concepts
       Minsup: Data Structure containing the supports of the different levels of
         abstraction.
       CT: Constraint defined by the user
1. Begin // Main Procedure
2. For (l = 1; l=< max_level; l++) do
3. {L [l, 1] = get_1_itemsets (T, l);
4.     For (k=2; L [l, k-1]! = null; k++) do
5.     {C [l, k] = get_Candidate_Set (L [l, k-1]);
6.      For each transaction t in T do
7.      {C_t = get_Subsets (C [l, K], t);
8.         For each   candidate c in C_t do  c.support++;
9.   }
10.     L [l, k] = {c Є C [l, k] | c.support >= minsup[l]};
```

```
11.        L [l, k] CT = satisfy_Constraint (L [l, k], CT); // the set of frequent itemsets
//satisfying the constraint CT
12.    }
13. LL[l]CT = Uk L [l, k]CT   ; // the set of all itemsets of the level l, satisfying the
//constraint CT
14. End;
```

```
// Pseudo-code of the function satisfy_Constraint:
1. Function satisfy_Constraint (L [l, k], CT)
2. Begin
3. For each itemset it Є L [l, k] do
4. {
5.      If (satisfy_Constraint_Itemset (it, CT)) Then
6.      {L [l, k]CT = L [l, k]CT U {it} ;}
7.}
8.  Return L [l, k]CT  ;
9. End;
```

```
// Pseudo-Code of function satisfy_Constraint_Itemset:
1. Function satisfy_Constraint_Itemset (it, CT)
2. Begin
3. Satisfied = false;
```
4. For ($c_i$ Є CT; ((satisfied= false) && (i=< n)); i++) do // $c_i$ **means one of the**
//**conjunctions of constraint CT**
```
5. {satisfied = true;
```
6. For ($e_{ij}$ Є $c_i$; ((j=< $n_i$) && (satisfied = true)); j++) do // $e_{ij}$ **means a literal, consisting**
//**of an item and, if necessary, a sign of negation.**
```
7. {lij = item (eij);
8.      If (lij = eji) then
9.      {    if ((NOT (lij Є it)) && (Not Exists (Descendant (lij) Є it))) then
10.          {satisfied = false ;}
11.     Else
12.     {    if ((lij Є it) && Exists (Descendant (lij) Є it)) then
13.          {satisfied = false ;}
14.}
15.}
16.}
17. Return satisfied;
18. End;
```

In conclusion, here are some comments for better clarification of the previous algorithm:
- The function *satisfy_Constraint*, called at line 11 of the main procedure, filter a set of frequent k-itemsets at a level l (any l, k), to get out a subset noted L [l, k]$^{CT}$ whose itemsets satisfy the constraint CT.

- The function *satisfy_Constraint* uses another function that is named *satisfy_Constraint_Itemset* (See line 5 of the function *satisfy_Constraint*). This function checks whether an itemset satisfies a constraint or not. It implements the principle of interpretation of constraints.

- The function *satisfy_Constraint_Itemset* itself uses a function called item (See line 7 of the *satisfy_Constraint_Itemset* function), with the parameter $e_{ij}$. This function returns the item involved in the element eij, after removal of the sign of negation.

- The function Descendant (See lines 9 and 12 of the *satisfy_Constraint_Itemset* function) returns all descendants of an item, given as input parameter.

**Illustration:**

For reasons of simplification, only a subset of the whole running example of the scenario1 is presented in what follows, based on the hierarchy of concepts and the database of Figures 2 and 3. CT is the constraint defined by the user:

$$CT = ((NON (3**)) AND (11*)) OR (2**)$$

A value of minimum support is assigned for each level, as the following table:

**Table 3: Illustration of the first algorithm for extracting multi-level frequent itemsets under constraints: Scenario1.**

| level | Support |
|-------|---------|
| 1 | 5 |
| 2 | 4 |
| 3 | 3 |

**Level 2:** *Minsup = 4*

| Itemsets | Support | | Itemsets | Support |
|----------|---------|---|----------|---------|
| {11*, 12*} | 4 | | ~~{12*, 31*}~~ | ~~2~~ |
| {11*, 21*} | 7 | | {12*, 32*} | 5 |
| {11*, 31*} | 4 | | {21*, 31*} | 4 |
| {11*, 32*} | 7 | | {21*, 32*} | 8 |
| {12*, 21*} | 5 | | ~~{31*, 32*}~~ | ~~3~~ |

L [2, 2]

| Itemsets | Support | | Itemsets | Support |
|----------|---------|---|----------|---------|
| {11*, 12*} | 4 | | {21*, 31*} | 4 |
| {11*, 21*} | 7 | | {21*, 32*} | 8 |
| {12*, 21*} | 5 | | | |

L [2, 2] $^{CT}$

| Itemsets | Support |  | Itemsets | Support |
|---|---|---|---|---|
| {11*, 12*, 21*} | 4 |  | {21*, 31*, 32*} | 4 |
| ~~{11*, 12*, 31*}~~ | ~~2~~ |  | {12*, 21*, 32*} | 5 |
| {11*, 12*, 32*} | 4 |  | {31*, 21*, 11*} | 4 |
| ~~{12*, 21*, 31*}~~ | ~~2~~ |  | {32*, 21*, 11*} | 7 |

L [2, 3]

| Itemsets | Support |  | Itemsets | Support |
|---|---|---|---|---|
| {11*, 12*, 21*} | 4 |  | {21*, 31*, 32*} | 4 |
| {31*, 21*, 11*} | 4 |  | {12*, 21*, 32*} | 5 |
| {32*, 21*, 11*} | 7 |  |  |  |

L [2, 3] $^{CT}$

An example illustrating a detailed implementation of scenario 1 is presented above, which helped generate the frequent itemsets satisfying the constraint CT and belonging to different levels of abstraction in the hierarchy of concepts in Table 1. Checking the validity compared to the constraint is done after the calculation of supports for all the itemsets. This induces the processing, for each pass, of the support of a large number of itemsets, which may not satisfy the constraint defined by the user. This operation consumes time and reduces the performance of the algorithm. The term pass is defined as the basic research stage of the k-frequent itemsets on a level of abstraction l (any l and k).

**3.3.2. Second Scenario: Approach "Test and Generate":** The approach presented in the scenario 1 is to verify the validity of the itemsets against the constraint after finding all frequent itemsets noted L [l, k]. This involves the calculation of the supports of candidate itemsets noted C [l, k], at each pass. Knowing that a large number of frequent itemsets do not satisfy the constraint defined by the user, another approach that avoids this loss of time is presented in this section. This approach consists in creating a filter on all candidate itemsets in each pass. This filter is designed to eliminate the itemsets which do not satisfy the constraint before calculating their supports. This ensures that the calculation of the support is applied only for itemsets that satisfy the user constraint.

Some examples can be extracted from the illustration of the section 3.3.1 such as:
- The set L [2, 1] contains 5 frequent itemsets; only 2 among them satisfy the constraint CT.
- The set L [3, 1] contains 7 frequent itemsets; only 3 among them satisfy the constraint CT.

The difference between the number of frequent itemsets and the itemsets that satisfy the constraint in the same pass is clear and remarkable. This difference increases certainly when dealing with real life large databases.

The application of the approach of scenario 2 improves the overall performance of the algorithm. However, it does not find all the frequent itemsets, but only a small part, because a k-itemset which does not satisfy a constraint, can participate to the generation of a (k +1)-Itemset that satisfies this constraint, based on A-priori.

The following examples are given to illustrate this approach:

- Consider all frequent 1-itemsets of level 2, L [2,1], the itemset {31*} is frequent but does not satisfy the constraint CT. Despite this, this itemset is used to generate the 2-itemset {21 *, 31 *} which is frequent and satisfies CT. Thus, if the itemset {31 *} has not been generated in L [2,1], we failed to find the itemset {21*, 31*)} in L [2,2]$^{CT}$.

- Consider all the frequent 2-itemsets of level 3, L [3,2], the itemset {111,321} is frequent but does not satisfy the constraint CT. However, this itemset is used to generate the 3-Itemset {111, 212, 321} which is frequent and satisfies the constraint CT.

The conclusion is that a frequent itemset which does not satisfy the constraint in one pass can contribute to the generation of frequent itemsets in the next pass. Indeed, the approach of scenario 2 leads to the omission of the generation of several frequent itemsets satisfying the constraint. Then this algorithm is incomplete and does not solve the problem and achieve the objectives. Another approach that is supposed to draw advantage from the real needs of the user is presented in the following section.

**3.3.3. Third Scenario: Pruning based Approach: Algorithm MLC-Prune:** This approach is based on a new method for introducing constraints by the user. This constraint is divided into two parts: the first is devoted to items that the user decides to remove from the mining process; the second is devoted to items that will be part of frequent itemsets.

**Modeling constraints:**
The constraint of the user is divided in two parts, called sub-constraints, the terminology of modelling of the constraints is defined in section 3.2:

- The first sub-constraint contains the literals or items, not covered by the user during the current search of frequent itemsets. This sub-constraint, noted CT_NEG (Negation), is modeled as follows:
  
  **CT_NEG = (NON $g_1$) AND (NON $g_2$) AND ................AND (NON $g_n$)**
  
  $g_i$ designate items or literals.

- The second sub-constraint contains literals or items that the user wishes to have in the itemsets to discover. This sub-constraint, noted CT_AFF (affirmation), is modeled in the form of disjunction of conjunctions:
  
  **CT_AFF = $c_1$ OR $c_2$ OR $c_3$ OR................OR $c_n$**
  
  $c_i$ is combination of items, with the following structure:
  
  **ci = ei1 AND ei2 AND ei3.........................AND eini**
  
  **eij** is an item or literal. The specificity of $e_{ij}$ in CT_AFF is that it can not contain a sign of negation. The use of negation sign is only done in the first sub-constraint CT_NEG.

**Example of constraint:**
The following example, denoted CT, is based on the database and the hierarchy of concepts of Figures 2 and 3:

**CT:** { CT_NEG = (NON (31*)) AND (NOT (112))
CT_AFF = (1** AND 21*) OR (3** AND 2**)

This technique is more suitable to the real needs of the user for the following reasons:

- The user focuses, in most cases, the search on a special axis (a well-defined subset).

- The rarity and inconsistency, in practice, of the constraints that contain an item that changes the sign from a conjunction to another.

**Example:**

25

$$CT = (1^{**} \text{ AND } (NON (3^{**}))) \text{ OR } (2^{**} \text{ AND } 3^{**})$$

- The potential improvement of performances.

**Principle:**
The search method of frequent multi-level itemsets under constraints with pre-pruning, presented in this scenario, proceed as follows:

- In a first step, it performs a pruning operation which consists in removing the items contained in the sub-constraint CT_NEG (Items removed by the user), from the database and the concept hierarchy. Recognizing that the elimination of one item of the concept hierarchy implies the elimination of all his descendants, i.e., a branch of the hierarchy, based on the principle of interpretation of the constraints presented in the section 3.2.

- In a second step, we proceed to searching frequent itemsets that satisfy the second sub-constraint CT_AFF, applying the principle of scenario 1, presented in the section 3.3.1.

The main contribution of this method is the pruning of the database and the concept hierarchy, which precedes the operation of discovering frequent itemsets. This reduces the itemsets lattice's size to run through for each level of abstraction. In addition to that, we verify the validity of frequent itemsets over the sub-constraint CT_AFF which does not contain literals with a sign of negation. This method operates on the definition of the constraints by the user and obliged him to divide it into two sub-constraints and make choices on items to eliminate from the search process.

**Table 4: Pseudo-code of the algorithm of mining frequent multi-level itemsets under constraints with pre-pruning: Scenario 3 : MLC_Prune s**

| |
|---|
| **Algorithm 2: Mining frequent multi-level itemsets under constraints with pruning: MLC-Prune** |
| Input:  **T**: Data base transactions |
|        **HC**: Hierarchy of concepts |
|        **Minsup**: Structure containing the supports of the different levels of abstraction |
|        **CT_NEG:** First Sub-constraint |
|        **CT_AFF:** Second Sub-constraint |
| 1. Begin // **Main Procedure** |
| 2. HC_Pruned = Pruning_Concept_Hierarchy (HC, CT_NEG); *// **pruning the //concept hierarchy using CT_NEG*** |
| 3. T_Pruned = Prunning_DB (T, CT_NEG); *// **pruning the database using //CT_NEG*** |
| 4. For (l = 1; l=< max_level; l++) do |
| 5. {L [l, 1] = get_1_itemsets (T_Pruned, l); |
| 6.     For (k=2; L [l, k-1]! = null; k++) do |
| 7.     {C [l, k] = get_Candidate_Set (L [l, k-1]); |
| 8.      For each transaction t in T_Pruned do |
| 9.     {$C_t$ = get_Subsets (C [l, K], t); |
| 10.       For each   candidate c in $C_t$ do  c.support++; |
| 11.     } |
| 12.     L [l, k] = {c Є C [l, k] | c.support >= minsup[l]}; |
| 13.     L [l, k]$^{CT}$ = Satisfy_Constraint_Pruning (L [l, k], CT_AFF); *// **the set of //frequent itemsets satisfying the constraint CT_AFF*** |
| 14.     } |

15. LL[l]$^{CT}$ = U$_k$ L [l, k]$^{CT}$  ; // **the set of all itemsets of the level l, satisfying the**
//**constraint CT**
16. end;

---

// function Pruning_Concept_Hierarchy :
1. Function Pruning_Concept_Hierarchy (HC, CT_NEG)
2. Begin
3. for each item gi in CT_NEG do
4. {HC_Pruned = HC - {gi, Descendant (gi)}; // **Elimination of the item gi and**  //**its descendants from HC**
5.}
6.  Return HC_Pruned;
7. End;

---

/ / function Prunning_DB:
1. Function Prunning_DB (HC, CT_NEG)
2. Begin
3. T_Pruned = T;
4. For each transaction t in T_Pruned do
5. {
6.  for each item g$_i$ in CT_NEG do
7. {
8.      T = t -  {g$_i$, Descendant (g$_i$)}; // **Elimination of gi item and its**
// **descendants from the current transaction**
9.}}
10.  Return T_Pruned;
11. End;

---

/ / function Satisfy_Constraint_Pruning:
1. Function Satisfy_Constraint_Pruning (L [l, k], CT_AFF)
2. Begin
3. For each itemset it Є L [l, k] do
4. {
5.      If (Satisfy_Constraint_ Itemset_Pruning (it, CT_AFF)) Then
6.      {L [l, k]$^{CT}$ = L [l, k]$^{CT}$ U {it} ;}
7.}
8.  Return L [l, k]$^{CT}$   ;
9. End;

---

// function Satisfy_Constraint_ Itemset_Pruning:
1. Function Satisfy_Constraint_ Itemset_Pruning (it, CT_AFF)
2. Begin
3. Satisfied = false;
4. For (c$_i$ Є CT_AFF; ((satisfied= false) && (i=< n)); i++) do // $c_i$   is one of the
//conjunctions of the sub-constraint CT_AFF
5. {satisfied = true;
6. For (e$_{ij}$ Є c$_i$; ((j=< n$_i$) && (satisfied = true)); j++) do // **e$_{ij}$ means a literal or an** //**item**
7. {
8.      If ((NOT (e$_{ij}$ Є it))&& (Not Exists ( Descendant (e$_{ij}$) Є it))) then
9.           {satisfied = false;   }
10.}}}
11. Return satisfied;

| 12. End; |
|---|

The following observations are presented to better clarify the algorithm:

- The main procedure of this algorithm begins with the pruning of the concept hierarchy and the database using the sub-constraint CT_NEG. This is done using Pruning_Concept_Hierarchy and Prunning_DB functions.

- Line 13 of the main procedure, calls the function Satisfy_Constraint_Pruning to identify the frequent itemsets that satisfy the sub-constraint CT_AFF.

- The Satisfy_Constraint_Pruning function uses another function called Satisfy_Constraint_Itemset_Pruning, which verifies the validity of an itemset in relation with the second sub-constraint CT_AFF.

**Example**

The following example illustrates the performance of our approach using the hierarchy of concepts of Figures 2 and 3. Given the constraint CT divided into two sub-constraints:

- CT_NEG = (NOT (3**))

- CT_AFF = ((11*) OR (2**)).

After the pruning of the hierarchy of concepts and the database, the following results in Figure 3 and Table 5 are obtained:

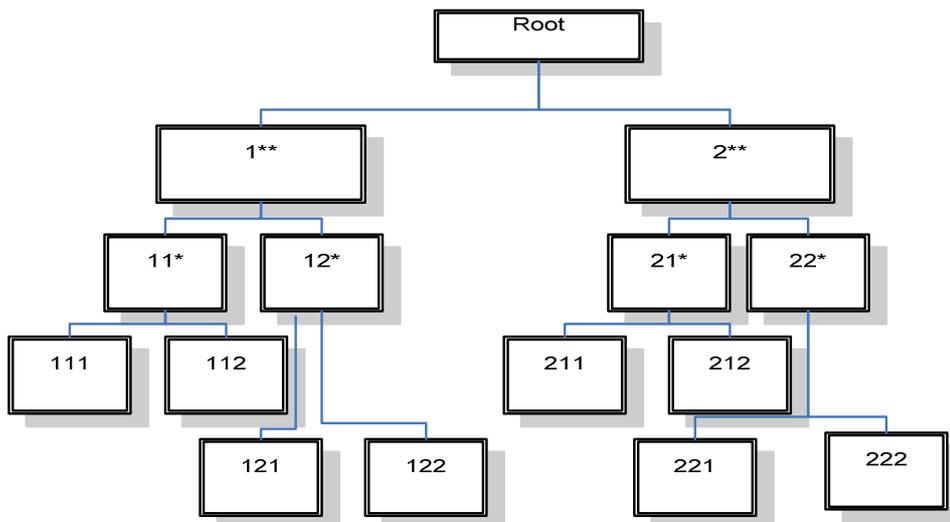

**Figure 3: Pruned Concept Hierarchy**

**Table 5: Pruned Database**

| TID | Items |
|---|---|
| 1 | 111, 212, 221 |
| 2 | 111, 122, 222, 212 |
| 3 | 122, 112, 212 |
| 4 | 212, 111, 122, 211 |
| 5 | 111, 211, 221 |

| 6 | 211, 121, 122 |
|---|---|
| 7 | 111, 212 |
| 8 | 212, 112, 122, 211 |

The reduced size of the concept hierarchy and the database after removal of the item (3**) and its descendants is noted. The following example details the execution illustration of this approach:

**Table 6: Illustration of the execution of the algorithm of mining frequent multi-level itemsets under constraints with Pre-Pruning: Third Scenario: MLC-Prune**

**Level 1 :** Minsup = 5

| Itemsets | Support |
|---|---|
| {1**} | 8 |
| {2**} | 8 |

L [1,1]

| Itemsets | Support |
|---|---|
| {2**} | 8 |

L [1,1] $^{CT}$

| Itemsets | Support |
|---|---|
| {1**, 2**} | 8 |

L [1,2]

| Itemsets | Support |
|---|---|
| {1**, 2**} | 8 |

L [1,2] $^{CT}$

**Level 2 :** Minsup = 4

| Itemsets | Support |
|---|---|
| {11*} | 7 |
| {12*} | 5 |
| {21*} | 8 |
| ~~{22*}~~ | ~~3~~ |

L [2, 1]

| Itemsets | Support |
|---|---|
| {11*} | 7 |
| {12*} | 5 |
| {21*} | 8 |

L [2, 1] $^{CT}$

| Itemsets | Support |
|---|---|
| {11*, 12*} | 4 |
| {11*, 21*} | 6 |
| {12*, 21*} | 5 |

L [2, 2]

| Itemsets | Support |
|---|---|
| {11*, 12*} | 4 |
| {11*, 21*} | 6 |
| {12*, 21*} | 5 |

L [2, 2] $^{CT}$

| Itemsets | Support |
|---|---|
| {11*, 12*, 21*} | 4 |

L [2, 3]

| Itemsets | Support |
|---|---|
| {11*, 12*, 21*} | 4 |

L [2, 3] $^{CT}$

**Level 3 :** Minsup = 3

| Itemsets | Support |
|---|---|

| Itemsets | Support |
|---|---|

| Itemsets | Support |
|---|---|

| {111} | 5 |
|---|---|
| ~~{112}~~ | ~~2~~ |
| ~~{121}~~ | ~~1~~ |
| {122} | 4 |

L [3, 1]

| {211} | 4 |
|---|---|
| {212} | 5 |
| ~~{221}~~ | ~~1~~ |
| ~~{222}~~ | ~~1~~ |

| {111} | 5 |
|---|---|
| {211} | 4 |
| {212} | 5 |

L [3, 1] $^{CT}$

| Itemsets | Support |
|---|---|
| ~~{111, 122}~~ | ~~2~~ |
| ~~{111, 211}~~ | ~~2~~ |
| ~~{111, 212}~~ | ~~2~~ |

L [3, 2]

| Itemsets | Support |
|---|---|
| ~~{122, 211}~~ | ~~2~~ |
| {122, 212} | 4 |
| ~~{211, 212}~~ | ~~2~~ |

| Itemsets | Support |
|---|---|
| {122, 212} | 4 |

L [3, 2] $^{CT}$

This running example shows clearly that the number of itemsets generated and analyzed was significantly reduced compared to the approach of the first scenario. This is due to the pruning phase, which has eliminated the item (3**) and its descendants in the database and the concept hierarchy using the sub-constraint CT_NEG.

**Rationale of the pruning based approach**

The approach proposed in this scenario comes in the goal of optimizing mining process of multi-level frequent itemsets under constraints. This approach requires that the user divides his constraint into two sub-constraints: CT_NEG and CT_AFF. CT_NEG includes the items that the user wishes to eliminate from the search process. CT_NEG is used to achieve a pruning operation of the database and the concept hierarchy. This pruning reduces the size of the trellis of itemsets for each level of abstraction and then the time reserved for the calculation of supports. In addition to that, transactions in the pruned database are smaller which improves the performance of the database scan. Moreover, the reduction in the number of discovered itemsets can improve their interpretation by the user. The technique used for modeling constraints of this approach allows the user to define its objectives in terms of the content of the discovered patterns. The effectiveness of this algorithm is validated when treating real life large databases.

**3.3.4. Experimentations:** In order to study the effectiveness of the approaches suggested for the resolution of the problem of mining multi-level frequent itemsets under constraints, a series of experiments were carried out. The algorithms related to approaches of scenarios 1 and 3, respectively, proposed in sections 3.3.1 and 3.3.2 were implemented. A generation of a database (With several sizes: 3000, 4000, 5000 and 6000 transactions with average of 8 items per transaction) and a concept hierarchy (With several sizes: 10, 30, 40 and 50 roots, e.g., items of level 1) on its items, was performed in order to experiment our algorithms. All experiments were performed under identical technical conditions. The implementation of the algorithms was carried out in following environments: Oracle *JDeveloper* to implement the algorithms, and an Oracle 10g server (*Sun Sparc E450, 1 GB RAM, OS: Unix Solaris 10*), for the database.

In a first experiment, several values of *minsup* (support threshold) were affected for the different levels (Level 1: 30%, Level 2: 20%, level 3: 10%). the size of the database was changed several times in order to study the impact of this change on the performance of scenarios 1 and 3. It should be noted that the constraints used for both scenarios are semantically equivalent. The results of this experiment are showed in Figure 4 below.

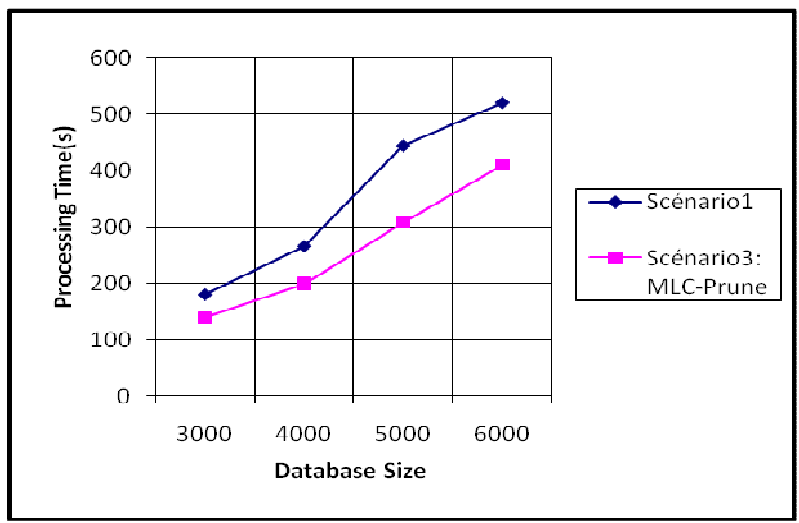

**Figure 4: Experimentation 1: Comparison of the performances of first and third scenario, depending on the size of the database**

The processing time of the third scenario is lower than the first scenario. This is due to the pruning step, in the third scenario, which reduces the number of itemsets analyzed and the complexity of the generation of candidates at each pass. The difference of execution time between the two algorithms increases with the increase in the size of the database.

In the second experimentation, a fixed size for the database has been set and we tried to study the impact of changes in the value *minsup* (support threshold) and to study the behavior of the algorithms of the first and third scenario. In this experiment, we assigned the same value of *minsup* for all levels of abstraction. The constraints are semantically equivalent. The results of this experiment are showed in Figure 5.

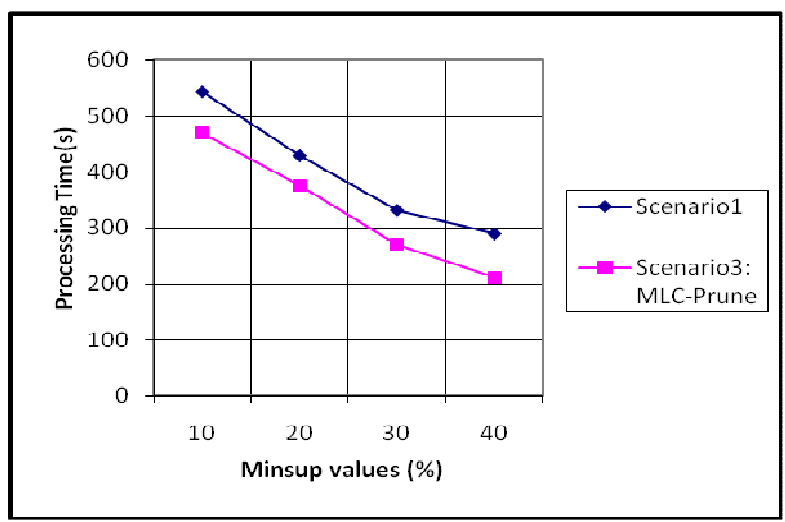

**Figure 5: Experimentation 2: Comparison of the first and third scenario depending on the values of minimum support.**

Similarly to the results of the first experimentation, the third scenario is more efficient than the first scenario. Performances of both algorithms better with higher support thresholds.

31

In the third experimentation, we studied the behavior of our algorithms, Scenario1: Basic and Scenario 3: MLC-Prune, under modification of the number of roots of the concept hierarchy, e.g., the number of items of level 1. We increased the number of roots from 10 to 50. As shown on figure 6, the processing time increases by increasing the number of roots. MLC-Prune is more efficient then the basic algorithm (Scenario1). The reason is that as the number of roots increases, the pruning step, implemented in MLC-Prune Algorithm, will have more interest and reduces significantly the itemsets lattice analyzed for each level of abstraction. Then, MLC-Prune treats always a smaller concept hierarchy and database. Furthermore, the constraints defined for algorithms processing handle with a high number of items, which harden the operation of checking itemsets validity and give more effectiveness to the pruning operation.

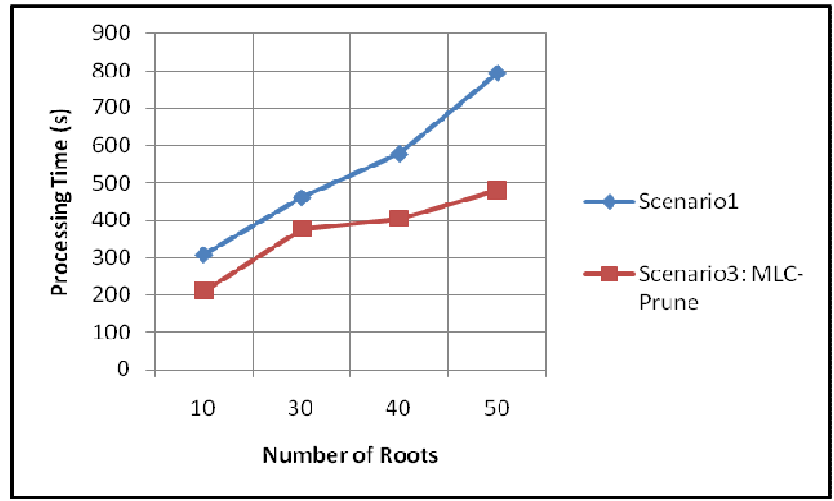

**Figure 6: Experimentation 3: Comparison of the first and third scenario (MLC-Prune) Depending on the number of roots of the Concept Hierarchy**

The fourth experimentation, whose results are shown in Figure7, presents a comparison between our algorithms and other algorithms of the literature, ML-T1 and ML-T2 proposed in [9]. ML-T2 and ML-T1 were implemented because they implement the same mining strategy with our algorithms. The number of database transactions was progressively increased and we executed all the algorithms in typically identical technical conditions. Results in Figure7 show that MLC-Prune Algorithm is the most efficient. This is heroically caused by the pruning step based on a very rich constraint, eliminating many branches of the concept hierarchy. The optimization technique implemented in ML-T2 algorithm which consists in pruning the database by eliminating all non frequent 1-itemsets of level 1 and their descendants wasn't very efficient as we assigned low support threshold for level 1 processing in ML-T2. In addition to that, this optimization technique may have more interest when handling more huge databases.

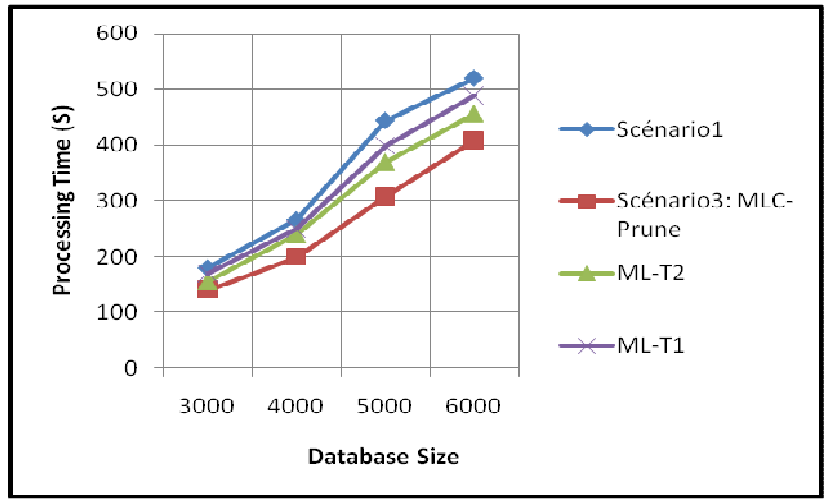

**Figure 7: Experimentation 4: Comparison of the first and third scenario (MLC-Prune) with ML-T2 and ML-T1 algorithms**

The results of the experimentations have confirmed our expectations on the theoretical level and have demonstrated the feasibility of our approaches.

## 4. Conclusion

In this paper, we introduced the problem of mining multiple level frequent itemsets under constraints, which allow the user to control the mining process and especially the existence of items into itemsets. We proposed a technique for modeling existence constraints in the context of use of concept hierarchies. Then, three algorithms were developed and studied to resolve this problem: The algorithm Basic (Scenario 1), the Test and Generate algorithm (Scenario 2) and the pruning based algorithm (Scenario 3). It is to note that it was proved that the algorithm of the scenario 2 is not complete and leads to the omission of a high number of frequent itemsets. Several Experimentations were performed in order to validate the algorithms we proposed in this paper and to study their behavior depending on some parameters such as database size and minimum support value. We have also compared our algorithms to other algorithms of the literature.

This work will be completed in our research group, by the design and implementation of a SQL like language that allows the expert to specify the minimum support for each level of the concept hierarchy and specify constraints; in addition to the introduction of other quality measures: confidence, lift, Loevinger, etc.

**Authors**

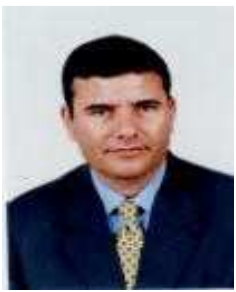
Mohamed Salah **GOUIDER** is associate professor at the University of Tunis – Tunisia. He now has thirty years experience in the field of databases and in recent years in the field of Data Warehouse and Data Mining. He earned his doctorate at the Faculty of Science - University of Nice - France in 1983. He is currently a consultant top management in several public and private enterprises. He has extensive experience in several countries: France, Tunisia, Kuwait, Qatar and Benin. His recent research is focused mainly in the extraction of knowledge from data in the medical and financial, as well as the optimization algorithms of the Data Warehouse and Data Mining.

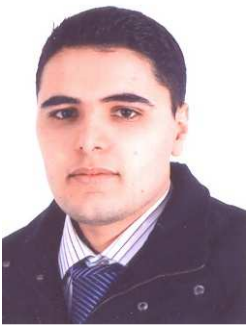
Amine **FARHAT**, Master of Science in Computer Science, is researcher and Ph.D. Student at the BESTMOD Laboratory, University of Tunis. He is assistant professor in Computer Science in the ISG of Tunis. He is also a senior computer science analyst Engineer, Head of Software Development and Decision Support Systems Projects, in a public Office. His research Interests are mainly Knowledge Discovery in Databases, Data mining algorithms, Artificial Intelligence and Data warehousing.